\documentclass{evn2024}
\usepackage{graphicx}
\usepackage{url}
\begin{document}
\raggedbottom
   \title{New insights on supernova remnants and HII regions in M82}

   \author{D. Williams-Baldwin\inst{1},
          G. Lucatelli\inst{1},
          T.W.B. Muxlow\inst{1},
          R.J. Beswick\inst{1},
          S.W. Shungube\inst{2},\\
          R. Lumpkin-Robbins\inst{1},
          M.K. Argo\inst{3},
          D.M. Fenech\inst{4},
          N. Kimani\inst{5},
            \and
          J. Radcliffe\inst{2,1}
          }

   \institute{Jodrell Bank Centre for Astrophysics, School of Physics and Astronomy, The University of Manchester, Manchester M13 9PL, UK
        \and 
   Department of Physics, University of Pretoria, Private Bag X20, Pretoria 0028, South Africa
        \and
   Jeremiah Horrocks Institute, University of Central Lancashire, Preston, Lancashire, PR1 2HE, UK
        \and
    SKAO, Jodrell Bank, Lower Withington, Macclesfield, SK11 9FT, UK
        \and
    Department of Physics, Kenyatta University, PO Box 43844, 00100 Nairobi, Kenya}

   \abstract{
   The nearby (d=3.6 Mpc) starburst galaxy M82 has been studied for several decades by very long baseline interferometry (VLBI) networks such as \textit{e}-MERLIN and the European VLBI Network (EVN). The numerous supernova remnants (SNRs), \ion{H}{ii} regions and other exotic transients make it a perfect laboratory for studying stellar evolution and the interstellar medium (ISM). Its proximity provides a linear resolution of 17 pc/arcsec, enabling decadal-time-scale variability and morphology studies of the tens of compact radio sources. In this proceedings, we describe new techniques developed in the last ten years that provide deeper, more robust imaging, enable in-band spectral index mapping, and allow wider fields of view to be imaged to find new radio sources.
   }
    \authorrunning{Williams-Baldwin et al. 2024}
   \maketitle
   
%

\section{Introduction}

Messier 82 (M82) is the perfect laboratory for studying stellar evolution, owing to its proximity ($d=3.6\,\mathrm{Mpc}$, \cite{Fenech2008}), its high star formation (SF) rate and consequently high supernova (SN) rate of 1 every 10-20 years (\cite{Fenech2010}). Optically, the emission is highly absorbed, necessitating high-resolution VLBI radio studies to infer the SF and SN rates. Over the past three decades, EVN and \textit{e}-MERLIN monitoring have studied the 10s of compact radio sources in the central kpc of M82. As the telescope sensitivity has improved, so has the number of sources observed, with more than 100 catalogued. The majority of the compact sources have been categorised as SN remnants (SNRs) or \ion{H}{ii} regions (see \cite{BeswickVLBIM822006,Fenech2008,Fenech2010,gendre13}) based on their morphology and in some cases their radio spectral index (S $\propto\nu^{\alpha}$, where $\alpha$ is the spectral index). Other than the common SNRs, \ion{H}{ii} regions, and the occasional new SN (e.g., SN2008iz, see \cite{Kimani2008iz}), M82 contains several `exotic' transient sources where long-term monitoring has been invaluable for understanding their origins. For example, the compact source 41.95+57.5 has been decreasing in flux by $\sim$8.8$\%$ per year, is likely responsible for M82's brighter single-dish flux in the 1960s and 1970s and could be the long-lived emission from a Gamma Ray Burst (\cite{Muxlow41.95}).
An `unusual' radio transient was discovered in 2009 \cite{TomtransientAtel}, which could be an extra-galactic analogue of SS433 (\cite{Joseph2011}, but see \cite{TomtransientLetter} also). While discoveries of transients are always possible in M82, monitoring the compact source population is necessary to understand how these objects interact with the interstellar medium, providing insight into the physics of stellar evolution. 


\section{2015 \textit{e}-MERLIN and 2022 EVN data}

\begin{figure*}
    \centering
    \includegraphics[width=0.23\linewidth]{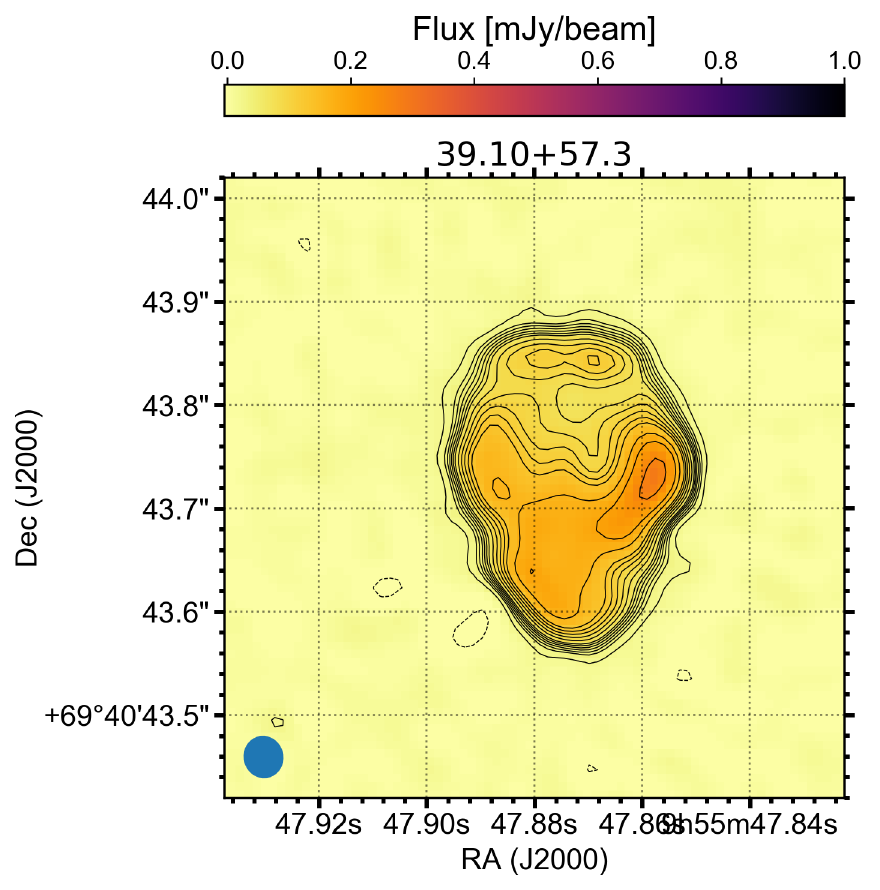}
    \includegraphics[width=0.23\linewidth]{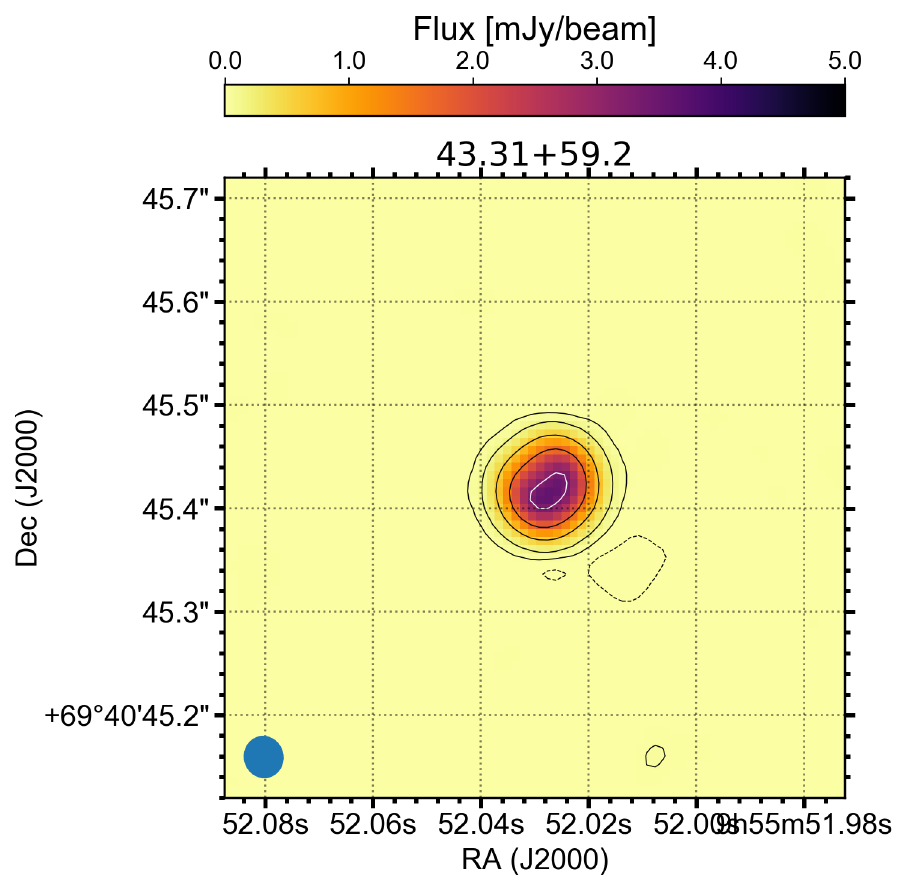}
    \includegraphics[width=0.23\linewidth]{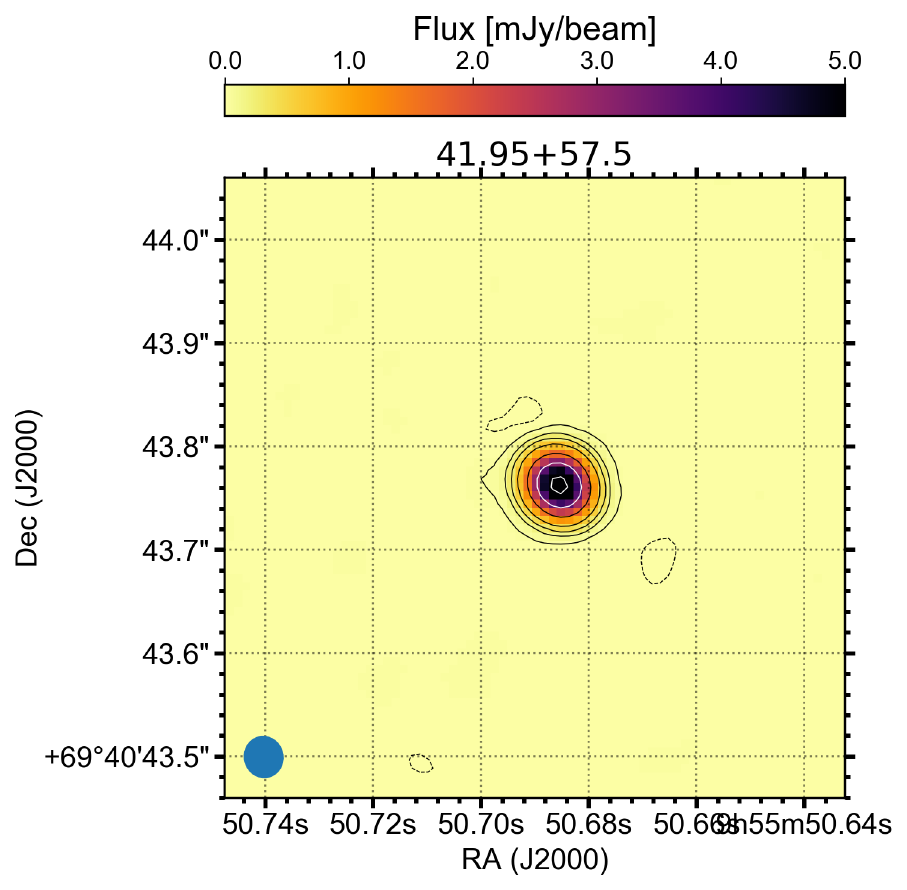}    
    \includegraphics[width=0.23\linewidth]{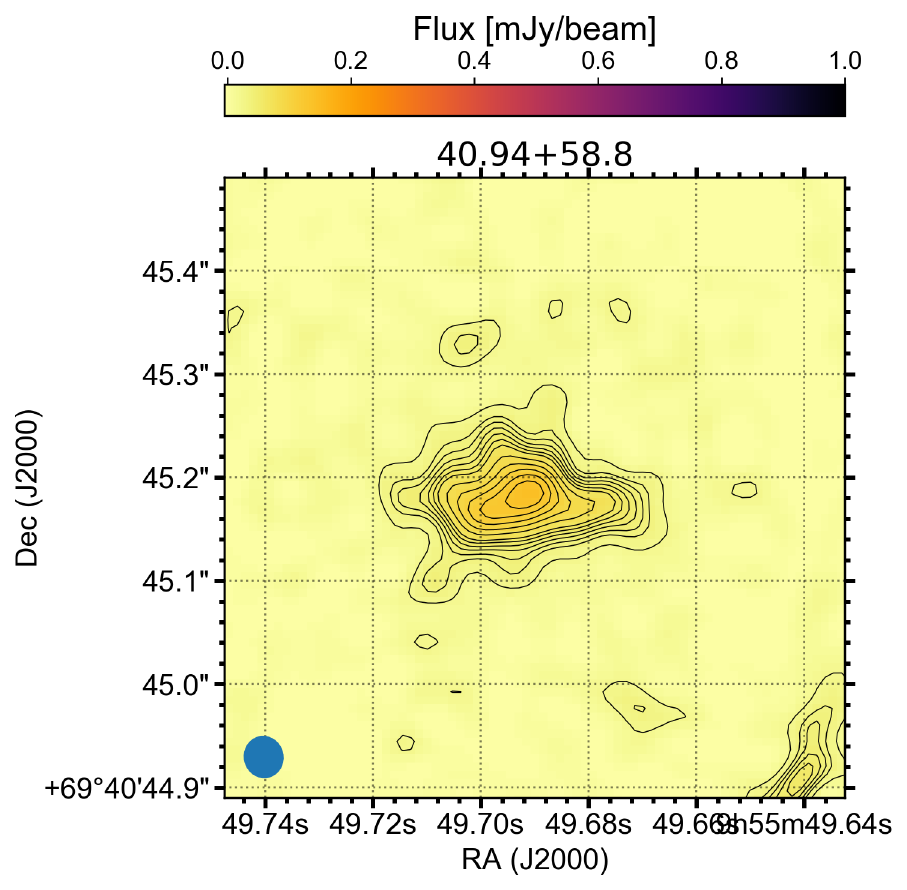} 
    \includegraphics[width=0.23\linewidth]{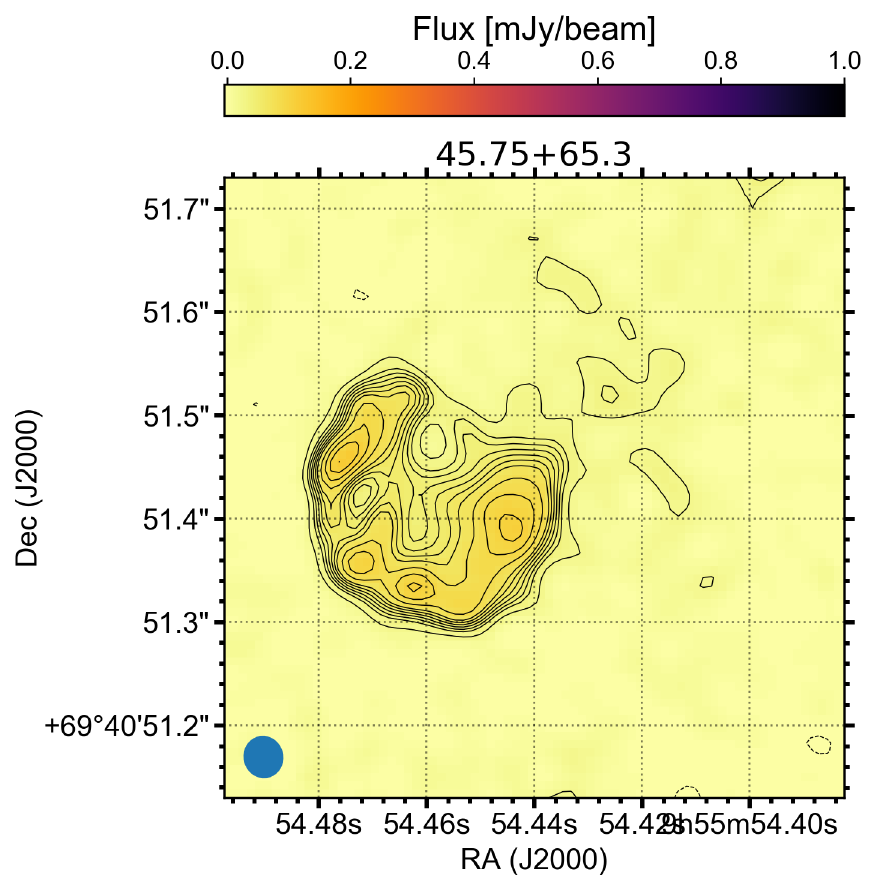}
    \includegraphics[width=0.23\linewidth]{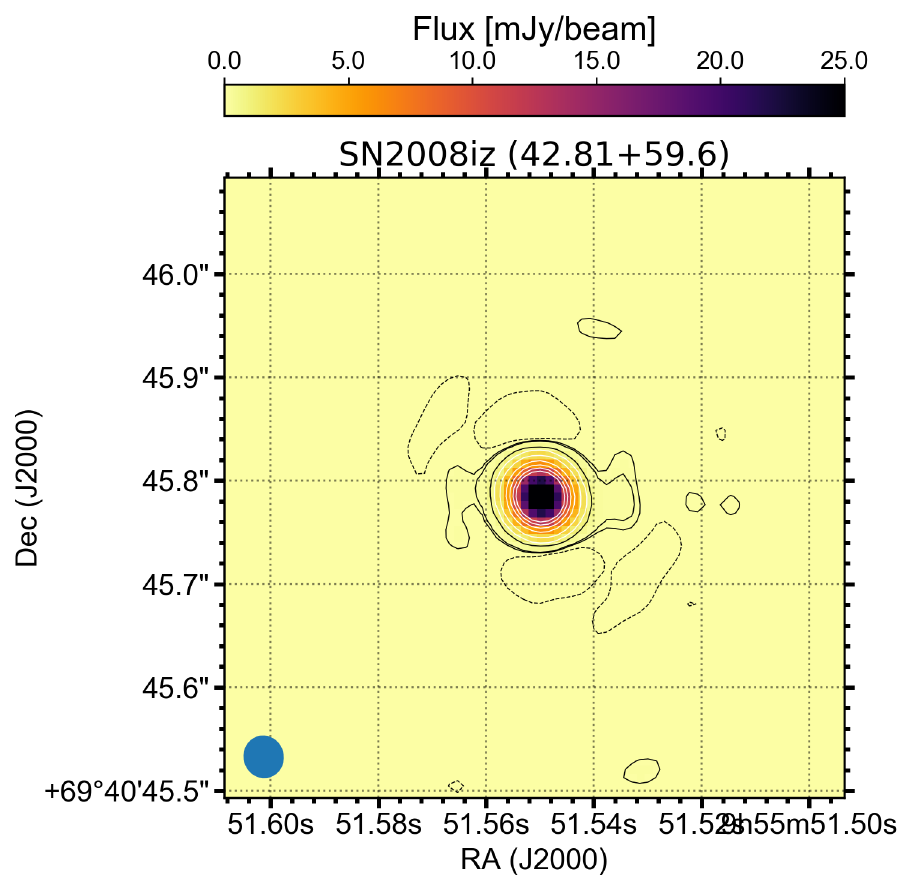}    \includegraphics[width=0.23\linewidth]{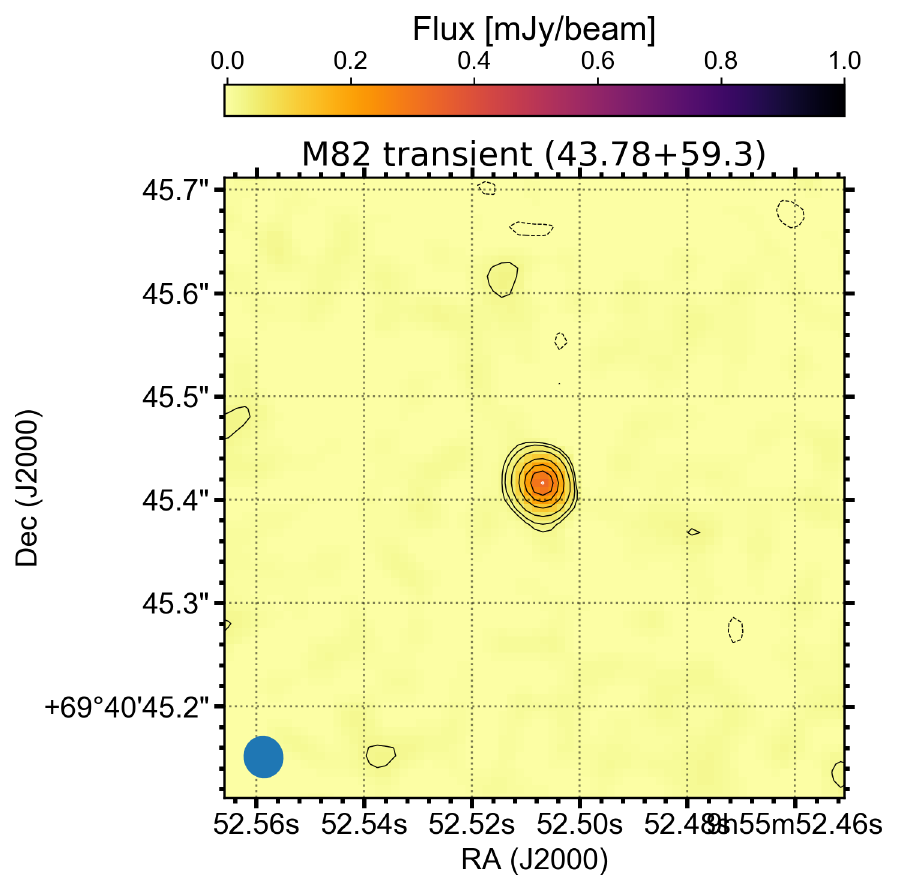}
    \includegraphics[width=0.23\linewidth]{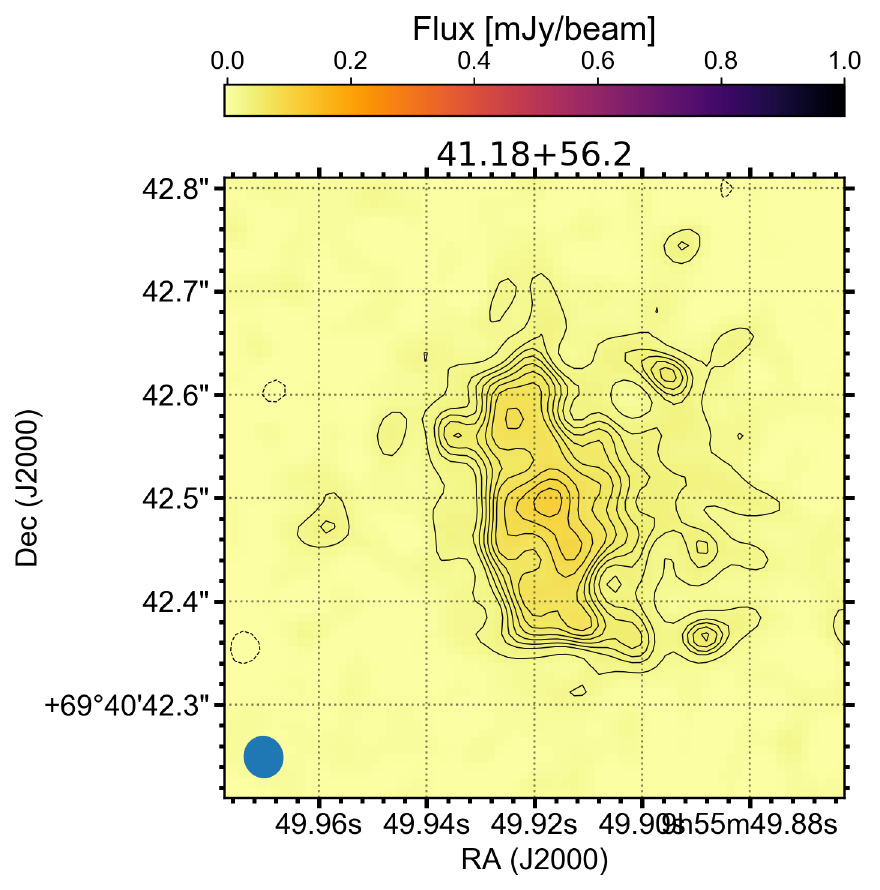}

    \caption{Eight sources obtained from the \textit{e}-MERLIN data presented here. The left column shows SNRs, where the expanding shells are clearly resolved by \textit{e}-MERLIN
    . The middle two columns show the compact younger supernova remnants 43.31+59.2 and SN2008iz
    and compact 'exotic' sources such as the "M82 transient" 
    and 41.95+57.5
    . The final column shows examples of faint low-surface brightness \ion{H}{ii} regions in M82, highlighting the need for sensitivity to detect these objects (see also Fig~\ref{fig:comp})
    . N.B. the contour levels are not identical between the images but instead highlight the main emission areas.}
    \label{fig:ims}
\end{figure*}

In this work, we present \textit{e}-MERLIN data taken in 2015 and preliminary results from European VLBI Network (EVN) data obtained in 2022. The \textit{e}-MERLIN data were obtained from 2015 May 28 to 2015 May 31, resulting in over 40 hours of on-source time. Observations were performed in two 512 MHz bands, centred at 5005 and 6202\,MHz, to provide a wider spectral index lever arm. The data were processed using the \textit{e}-MERLIN \textsc{CASA} Pipeline (\cite{eMCP}). \texttt{wsclean} (\cite{offringa-wsclean-2014,offringa-wsclean-2017}) software was used to image and self-calibrate the \textit{e}-MERLIN data. 
We used a `thresholded masking' method, 
to gradually include more of the radio sources in the model\footnote{We used a modified version of a thresholded mask code written by Ian Heywood for \texttt{OXKAT}: \url{https://github.com/IanHeywood/oxkat/blob/master/tools/make_threshold_mask.py}. The mask files were inspected after each new one was generated to ensure only real components were included in the imaging mask}, taking care not to include imaging artefacts. In \texttt{wsclean}, we 
included the \textit{e}-MERLIN primary beam\footnote{The primary beam models were made by Nicholas Wrigley and added to a \texttt{wsclean} container by Bob Watson. The container is available on request.} and multi-scale cleaning 
to fit the spectrum across the eight non-contiguous spectral windows. The final image achieved an r.m.s noise level of 6.5$\mu$Jy/beam. The image quality close to the bright SN2008iz is poorer, but further work is underway to mitigate its effect on the wider image. Some of the sources are presented in Figure~\ref{fig:ims}.

We also present a preliminary analysis from EVN data observed in June 2022 (Project code: EW029, PI: Williams). These observations were made in both L and C bands with 18 hours on the source in both bands, using EVN+\textit{e}-MERLIN to sample a wider range of spatial scales necessary for detecting faint diffuse emission in SNRs. To sample the full central kpc of M82, we used multiple correlation centres tessellated across this region. We included five additional correlation centres for off-nuclear sources of interest that had been picked up in archival widefield imaging of the M82 field in the LeMMINGs survey  (\cite{BaldiLeMMINGs2}). The EVN data were processed using \texttt{VPIPE} (\cite{jack_radcliffe_2024_11108171}), a \textsc{CASA} pipeline designed for calibrating VLBI data. \texttt{VPIPE} is particularly useful as it performs primary beam correction for EVN data, and casts the solutions from the pointing centre to the other correlation centres used in this project. The full dataset is still being analysed, but we show a preliminary L band image of SNR 43.31+59.2 in Figure~\ref{fig:SNR}.

\section{Results and Discussion}

\begin{figure*}
    \centering
    \includegraphics[width=0.54\linewidth]{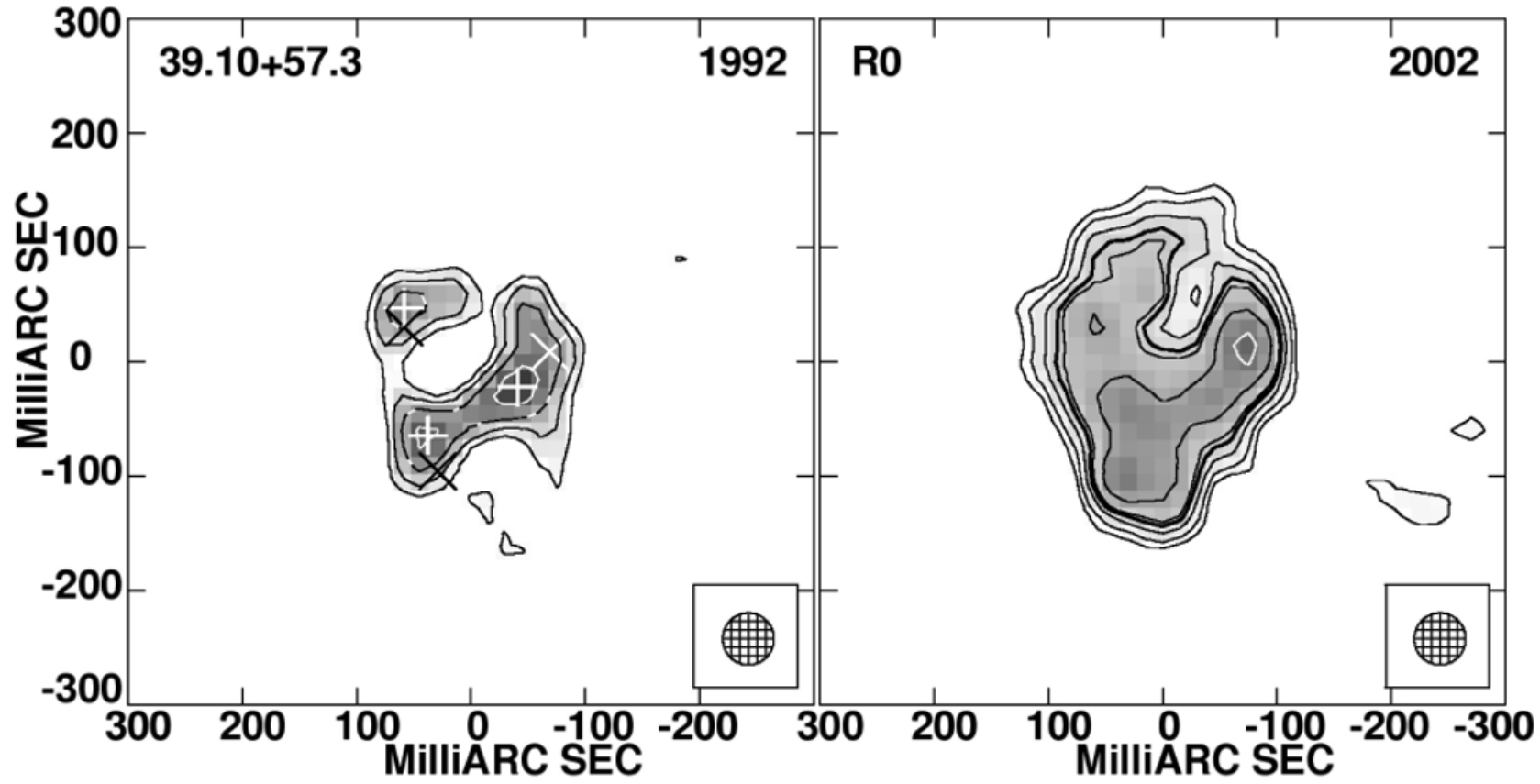}
    \includegraphics[width=0.33\linewidth]{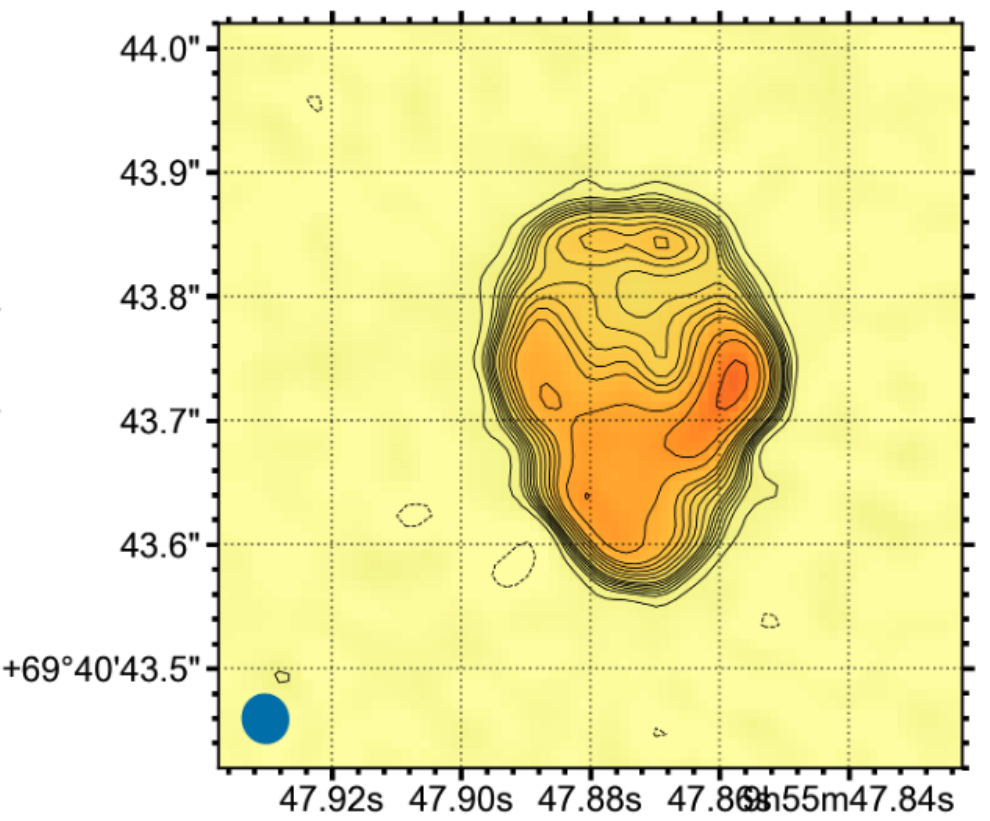}
    \includegraphics[width=0.54\linewidth]{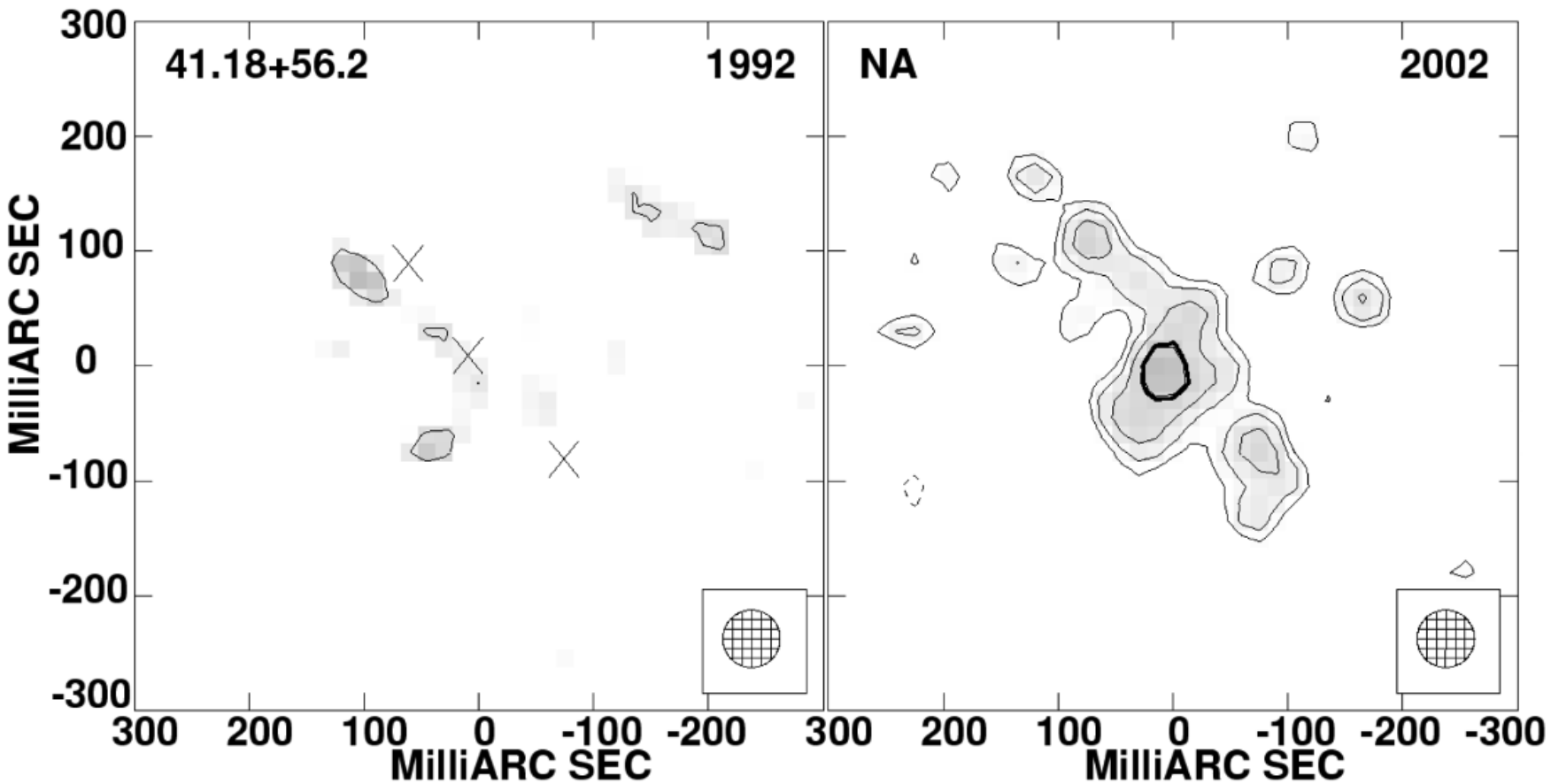}
    \includegraphics[width=0.33\linewidth]{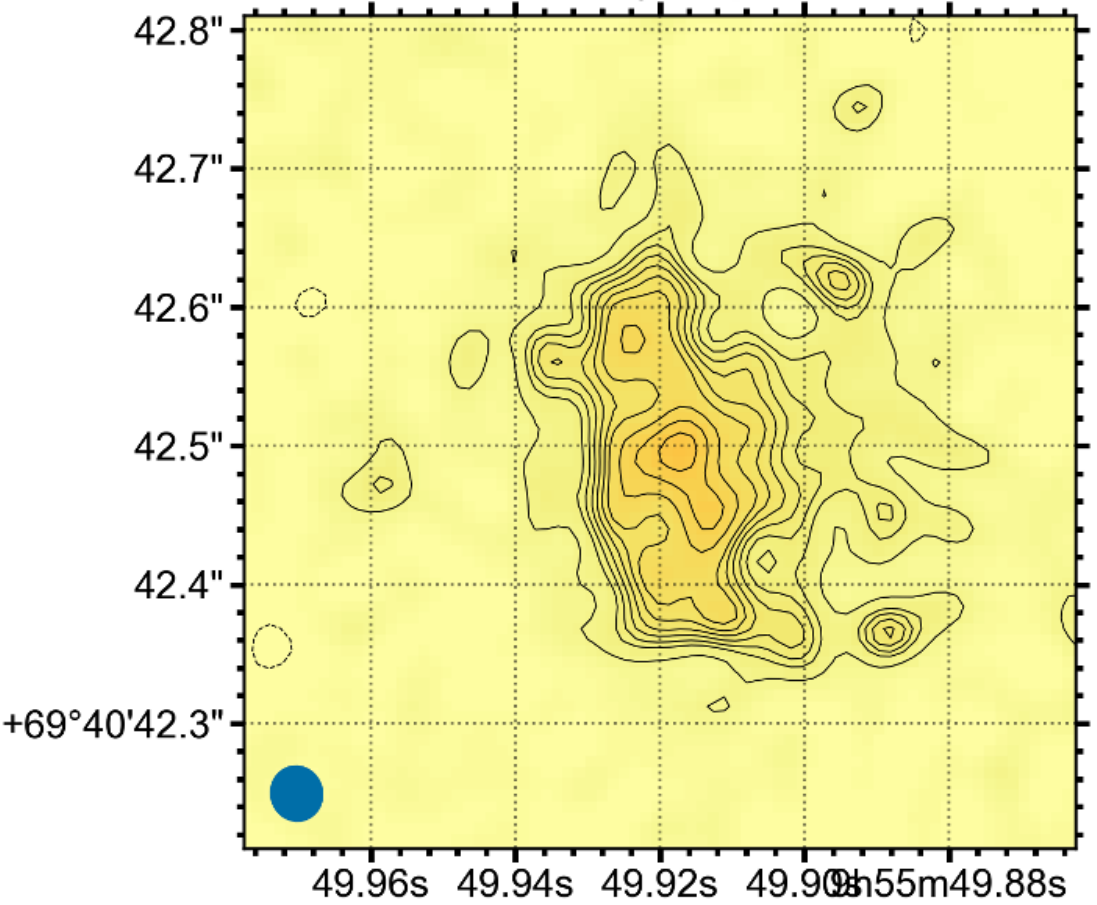}\\
    \caption{Improvement in MERLIN/\textit{e}-MERLIN image quality over 3 decades: the top row represents a classical SNR, 39.10+57.3, and the bottom row represents a HII region, 41.18+56.2. From left-to-right, the panels shown are: \textit{left:} 1992 MERLIN image, obtained from \cite{Fenech2008}, \textit{middle:} 2002 MERLIN image, obtained from \cite{Fenech2008} and \textit{right:} \textit{e}-MERLIN image presented from the data in this work. }
    \label{fig:comp}
\end{figure*}

\subsection{Improvement in imaging in 30 years}

Figure~\ref{fig:comp} shows the improvement in image fidelity and sensitivity over the last 30 years with MERLIN (the predecessor of \textit{e}-MERLIN), and the 2013 data presented here. The increasing improvement in sensitivity is evident, with lower surface brightness features and diffuse components becoming easier to detect in the most recent \textit{e}-MERLIN data. While the SNRs appear mostly the same as they did in the 2002 maps (\cite{Fenech2008}), the additional sensitivity is of critical importance for recovering the total flux of \ion{H}{ii} regions in particular: the \ion{H}{ii} regions are much larger than previously observed and consequently have larger total fluxes. The new \textit{e}-MERLIN map has also discovered ten new sources of radio emission at levels of $\sim$50$\mu$Jy/beam. Of the five additional off-nuclear sources found in archival \textit{e}-MERLIN data, 3/5 were detected in the 2022 EVN observations, showing that \textit{e}-MERLIN's sensitivity is ideal for finding uncataloged sources in M82.

\begin{figure}
    \centering
    \includegraphics[width=0.8\linewidth]{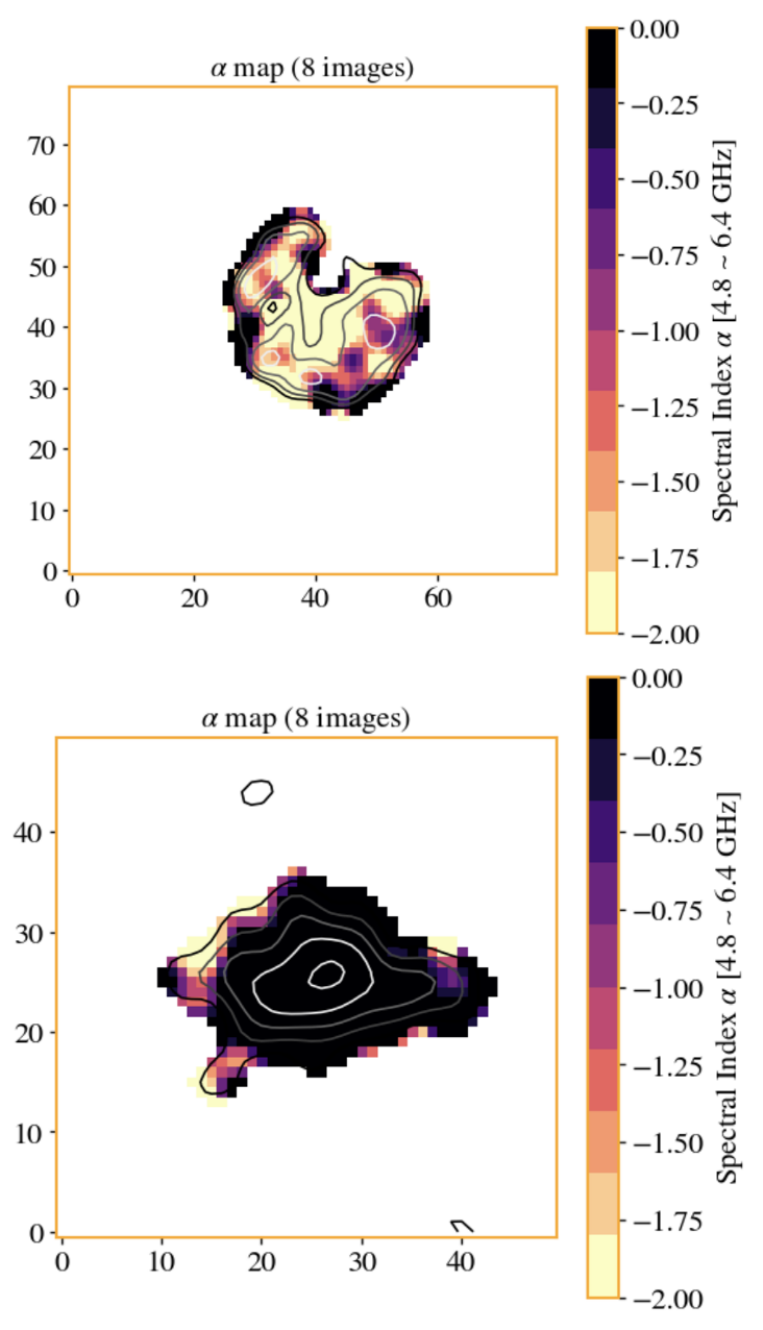}
    \caption{\textit{Top:} Spectral index colour map of SNR 45.75+65.3 and \textit{bottom:} \ion{H}{ii} region 40.94+58.8, both made using the \texttt{morphen} code (\cite{2024ascl.soft05009L}). Contours are plotted on top to show the regions of brighter flux. The SNR shows flatter spectra at the points where the flux is brighter (see Figure\ref{fig:ims} bottom left panel), whereas the \ion{H}{ii} region is consistent with a flat or slightly inverted spectrum across the entire source.}
    \label{fig:specind}
\end{figure}

\subsection{In-band Spectral Index maps}
We made pixel-by-pixel spectral index maps of the sources, using the \texttt{morphen} code (\cite{2024ascl.soft05009L})\footnote{The \textit{morphen} code was originally designed for decomposing emission from different mechanisms in nearly ULIRGs (see \cite{Lucatelli}), but can be applied here in the simplistic case of providing a spectral index map. Geferson Lucatelli provided a version of the code for this purpose.}. To make the requisite maps, we remade our images using a common circular restoring beam. Using \texttt{morphen}, we fit the spectral index across the \textit{e}-MERLIN band and separate the SNRs - which tend to have steeper spectra and circular morphologies - from the \ion{H}{ii} regions - which have flat or inverted spectra and complex morphologies - for the first time using a single dataset (see Figure~\ref{fig:specind}). This method is a significant improvement on previous works, where radio fluxes may have been obtained with different arrays and at various times, which could lead to issues with sampling the same spatial scales or be affected by source variability (e.g., see \cite{Fenech2010}). The \texttt{morphen} code also calculates source sizes and computes fluxes, with a full analysis planned for an upcoming paper (Williams-Baldwin et al., in prep.)

\subsection{The inclusion of \textit{e}-MERLIN in EVN}

\begin{figure}
    \centering
    \includegraphics[width=0.8\linewidth]{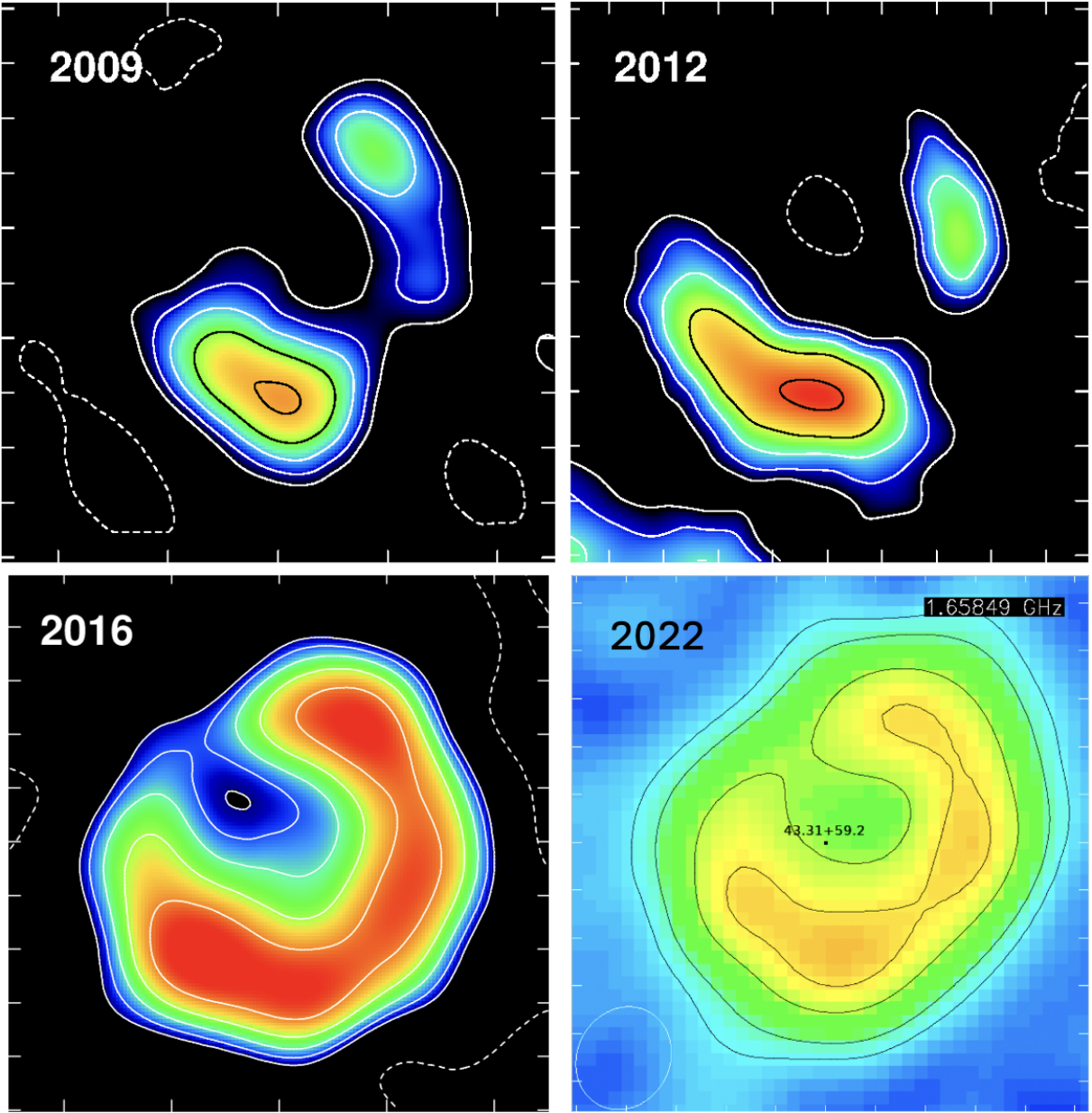}
    \caption{\textit{Top row:} C band EVN observations of SNR 43.31+59.2 in 2009 and 2012 without the \textit{e}-MERLIN array included. \textit{Bottom row:} 2016 C band (left) and 2022 L band (right) observations of the same source including \textit{e}-MERLIN. The diffuse flux is recovered only when including \textit{e}-MERLIN with the EVN, enabling longer-term study.}
    \label{fig:SNR}
\end{figure}

\textit{e}-MERLIN provides shorter baselines than the EVN and is sensitive to structures on 10s mas scales. This is critical for studying SNRs, as they will eventually become resolved by the EVN and undetectable. Combined EVN+\textit{e}-MERLIN observations provide the best of both: mas-scale resolution with EVN baselines and sensitivity to larger angular scales with \textit{e}-MERLIN. This is demonstrated in Figure~\ref{fig:SNR} for SNR 43.31+59.2, which is only slightly resolved in \textit{e}-MERLIN-only data (see Figure~\ref{fig:ims}). EVN-only data in 2009 and 2012 (Fenech et al., in prep) show that the 10s-mas-scale flux in this source to be resolved out, but the inclusion of \textit{e}-MERLIN in the 2016 C band and 2022 L band data presented here recovers much of the lost flux. 



\section{Conclusions}
We have presented new data analysis techniques to provide further insight on the compact source population found in M82. Not only have these new techniques enabled the detection of previously uncatalogued sources, but they have also shown that i) the \ion{H}{ii} regions in M82 are larger than previously thought with a greater total flux density, ii) in-band \textit{e}-MERLIN radio spectra are reliable for many of the SNRs and \ion{H}{ii} regions, with resolved spectral index maps a useful tool in diagnosing the source type, and iii) the inclusion of \textit{e}-MERLIN in the EVN is necessary for detecting the SNRs as they continue to expand and evolve as they will become resolved if the shorter spacings of \textit{e}-MERLIN are not included. Furthermore, wide-field \textit{e}-MERLIN observations can be used to find faint radio source emissions away from the central regions of galaxies, which can then be followed up with the EVN to identify the origin of these uncatalogued sources.

\begin{acknowledgements}
The European VLBI Network (www.evlbi.org) is a joint facility of independent European, African, Asian, and North American radio astronomy institutes. Scientific results from data presented in this publication are derived from the following EVN project code(s): EW029.
$e$-MERLIN is a National Facility operated by the University of Manchester at Jodrell Bank Observatory on behalf of STFC. 

\end{acknowledgements}

\bibliographystyle{apalike}


\bibliography{bib} 

\begin{thebibliography}{}

\bibitem[{Baldi} et~al., 2021]{BaldiLeMMINGs2}
{Baldi}, R.~D., {Williams}, D.~R.~A., {McHardy}, I.~M., {Beswick}, R.~J., {Brinks}, E., {Dullo}, B.~T., {Knapen}, J.~H., {Argo}, M.~K., {Aalto}, S., {Alberdi}, A., {Baan}, W.~A., {Bendo}, G.~J., {Corbel}, S., {Fenech}, D.~M., {Gallagher}, J.~S., {Green}, D.~A., {Kennicutt}, R.~C., {Kl{\"o}ckner}, H.~R., {K{\"o}rding}, E., {Maccarone}, T.~J., {Muxlow}, T.~W.~B., {Mundell}, C.~G., {Panessa}, F., {Peck}, A.~B., {P{\'e}rez-Torres}, M.~A., {Romero-Ca{\~n}izales}, C., {Saikia}, P., {Shankar}, F., {Spencer}, R.~E., {Stevens}, I.~R., {Varenius}, E., {Ward}, M.~J., {Yates}, J., and {Uttley}, P. (2021).
\newblock {LeMMINGs - II. The e-MERLIN legacy survey of nearby galaxies. The deepest radio view of the Palomar sample on parsec scale}.
\newblock {\em \mnras}, 500(4):4749--4767.

\bibitem[{Beswick} et~al., 2006]{BeswickVLBIM822006}
{Beswick}, R.~J., {Riley}, J.~D., {Marti-Vidal}, I., {Pedlar}, A., {Muxlow}, T.~W.~B., {McDonald}, A.~R., {Wills}, K.~A., {Fenech}, D., and {Argo}, M.~K. (2006).
\newblock {15 years of very long baseline interferometry observations of two compact radio sources in Messier 82}.
\newblock {\em \mnras}, 369(3):1221--1228.

\bibitem[{Fenech} et~al., 2010]{Fenech2010}
{Fenech}, D., {Beswick}, R., {Muxlow}, T.~W.~B., {Pedlar}, A., and {Argo}, M.~K. (2010).
\newblock {Wide-field Global VLBI and MERLIN combined monitoring of supernova remnants in M82}.
\newblock {\em \mnras}, 408(1):607--621.

\bibitem[{Fenech} et~al., 2008]{Fenech2008}
{Fenech}, D.~M., {Muxlow}, T.~W.~B., {Beswick}, R.~J., {Pedlar}, A., and {Argo}, M.~K. (2008).
\newblock {Deep MERLIN 5GHz radio imaging of supernova remnants in the M82 starburst}.
\newblock {\em \mnras}, 391:1384--1402.

\bibitem[{Gendre} et~al., 2013]{gendre13}
{Gendre}, M.~A., {Best}, P.~N., {Wall}, J.~V., and {Ker}, L.~M. (2013).
\newblock {The relation between morphology, accretion modes and environmental factors in local radio AGN}.
\newblock {\em \mnras}, 430:3086--3101.

\bibitem[{Joseph} et~al., 2011]{Joseph2011}
{Joseph}, T.~D., {Maccarone}, T.~J., and {Fender}, R.~P. (2011).
\newblock {The unusual radio transient in M82: an SS 433 analogue?}
\newblock {\em \mnras}, 415(1):L59--L63.

\bibitem[{Kimani} et~al., 2016]{Kimani2008iz}
{Kimani}, N., {Sendlinger}, K., {Brunthaler}, A., {Menten}, K.~M., {Mart{\'\i}-Vidal}, I., {Henkel}, C., {Falcke}, H., {Muxlow}, T.~W.~B., {Beswick}, R.~J., and {Bower}, G.~C. (2016).
\newblock {Radio evolution of supernova SN 2008iz in M 82}.
\newblock {\em \aap}, 593:A18.

\bibitem[{Lucatelli}, 2024]{2024ascl.soft05009L}
{Lucatelli}, G. (2024).
\newblock {morphen: Astronomical image analysis and processing functions}.
\newblock Astrophysics Source Code Library, record ascl:2405.009.

\bibitem[{Lucatelli} et~al., 2024]{Lucatelli}
{Lucatelli}, G., {Beswick}, R.~J., {Mold{\'o}n}, J., {P{\'e}rez-Torres}, M.~A., {Conway}, J.~E., {Alberdi}, A., {Romero-Ca{\~n}izales}, C., {Varenius}, E., {Kl{\"o}ckner}, H.-R., {Barcos-Mu{\~n}oz}, L., {Bondi}, M., {Garrington}, S.~T., {Aalto}, S., {Baan}, W.~A., and {Pihlstr{\"o}m}, Y.~M. (2024).
\newblock {The PARADIGM project I: a multiscale radio morphological analysis of local U/LIRGS}.
\newblock {\em \mnras}, 529(4):4468--4499.

\bibitem[{Moldon}, 2018]{eMCP}
{Moldon}, J. (2018).
\newblock {The new e-MERLIN CASA pipeline}.
\newblock In {\em 14th European VLBI Network Symposium \& Users Meeting (EVN 2018)}, page 152.

\bibitem[{Muxlow} et~al., 2010]{TomtransientLetter}
{Muxlow}, T.~W.~B., {Beswick}, R.~J., {Garrington}, S.~T., {Pedlar}, A., {Fenech}, D.~M., {Argo}, M.~K., {van Eymeren}, J., {Ward}, M., {Zezas}, A., and {Brunthaler}, A. (2010).
\newblock {Discovery of an unusual new radio source in the star-forming galaxy M82: faint supernova, supermassive black hole or an extragalactic microquasar?}
\newblock {\em \mnras}, 404(1):L109--L113.

\bibitem[{Muxlow} et~al., 2009]{TomtransientAtel}
{Muxlow}, T.~W.~B., {Beswick}, R.~J., {Pedlar}, A., {Fenech}, D., {Argo}, M.~K., {Ward}, M.~J., and {Zezas}, A. (2009).
\newblock {Discovery of a new transient radio source in the central region of M82}.
\newblock {\em The Astronomer's Telegram}, 2073:1.

\bibitem[{Muxlow} et~al., 2005]{Muxlow41.95}
{Muxlow}, T.~W.~B., {Pedlar}, A., {Beswick}, R.~J., {Argo}, M.~K., {O'Brien}, T.~J., {Fenech}, D., and {Trotman}, W. (2005).
\newblock {Is 41.95+575 in M82 actually an SNR? .}
\newblock {\em \memsai}, 76:586.

\bibitem[Offringa et~al., 2014]{offringa-wsclean-2014}
Offringa, A.~R., McKinley, B., Hurley-Walker, et~al. (2014).
\newblock {WSClean: an implementation of a fast, generic wide-field imager for radio astronomy}.
\newblock {\em MNRAS}, 444(1):606--619.

\bibitem[Offringa and Smirnov, 2017]{offringa-wsclean-2017}
Offringa, A.~R. and Smirnov, O. (2017).
\newblock {An optimized algorithm for multiscale wideband deconvolution of radio astronomical images}.
\newblock {\em MNRAS}, 471(1):301--316.

\bibitem[Radcliffe, 2024]{jack_radcliffe_2024_11108171}
Radcliffe, J. (2024).
\newblock jradcliffe5/vlbi\_pipeline: v1.1.

\end{thebibliography}
%
%
%
%
%
%
%
%
%
%
%

\end{document}